# Biochemical Mechanism of Resistance in Some *Brassica* Genotypes against *Lipaphis erysimi* (Kaltenbach) (Homoptera: Aphidiae)

**Sarwan Kumar* and M. K. Sangha**

Department of Plant Breeding and Genetics, Punjab Agricultural University, Ludhiana-141 004, India

Two years study was carried out during 2006-07 and 2007-08 crop seasons to study the response of different genotypes of oilseeds *Brassica* to *Lipaphis erysimi* (Kaltenbach) infestation both under field and screen house conditions and to find out the relationship of various biochemical constituents to aphid infestation. Among the various genotypes, the population of *L. erysimi* was significantly high on *Brassica rapa* variety brown *sarson* cv. BSH 1 and *B. rapa* var. yellow *sarson* cv. YST 151 in unprotected set i.e. 53.7 and 52.3 aphids/ plant, respectively. However, it was the lowest on *Eruca sativa* cv. T 27 (4.7 aphids/plant) followed by *B. carinata* cv. DLSC 2 (20.9 aphids/ plant) which suffered the least yield loss i.e. 5.79 and 10.59 per cent, respectively. Almost similar trend was observed in seedling mortality, which was the maximum in BSH 1 and YST 151, while no seedling mortality was observed in the case of T 27 during both the years of study. Analysis of various biochemical constituents revealed that glucosinolates, total phenols and ortho-dihydroxy phenols had inverse relationship with the aphid infestation. Higher amount of these biochemical constituents in T 27 and DLSC 2 was responsible for lower aphid infestation on these genotypes.

## INTRODUCTION

The genus *Brassica* comprises an important group of cultivated vegetable and oilseed crops and have great economic importance worldwide. The cultivated species of Brassica include cabbage, cauliflower, kale, broccoli, Brussels sprouts, turnip, swede, oilseed rape, various mustards and other leafy vegetables (Hong *et al.* 2008). As oilseed crop, rapeseed contributed 58.56 million metric tons out of total oilseed crop production of 446.97 million metric tons amounting 13.1% in the year 2010-11 (Source: Foreign Agricultural Service, Circular Series FOP 04–11, USDA). India is one of the leading producers including Canada, USA, EU, Australia and China. In India, under the name rapeseed and mustard three cruciferous members of *Brassica* species are cultivated; *B. juncea*

(Indian mustard or commonly called *rai*) being the chief oil-yielding crop, while three ecotypes of *B. rapa* ssp. *oleifera*, viz. brown *sarson*, yellow *sarson*, toria and *B. napus* are grown to a limited extent. Among the biotic stresses, the damage caused by aphids is a major constraint in the productivity of these crops. Among the three taxonomic variants that infest oilseed *Brassicas* in India, the turnip/mustard aphid, *Lipaphis erysimi* (Kaltenbach) (Homoptera: Aphididae), damages the crop ranging from 9 to 96 per cent in different agro-climatic conditions of the country (Singh and Sharma 2002) and is a major pest on oilseed *brassicas* in Indian sub-continent, while *Brevicoryne brassicae* (L.) has global presence with strong yield reducing impact, especially in vegetable *brassicas* (Kumar *et al.* 2011). Because of prolific breeding and short generation time, it multiplies very fast feeding exclusively on phloem sap. In the absence of control measures, it may lead to complete crop failure. For the management of this pest, farmers are dependent on the use of toxic insecticidal chemicals with systemic mode of action to a large extent as they find it the most easily available and effective method for pest management. They are little worried about the adverse effects of these insecticides *viz.* development of resistance to commonly used insecticides, pest resurgence, secondary pest outbreak(s) besides environmental pollution and pesticide residues in oil and cake and, consequently, in the food chain (Singh 2001, Singh and Sharma 2002).

Therefore, it becomes imperative that available pest management tactics should be such that provide effective and economical control of the pest without any adverse effect on the environment. In this context, the host plant resistance holds a promise. However, development of an insect resistant cultivar is a long process. The first step in the development of an insect resistant cultivar is the precise knowledge of sources of resistance (Stoner and Shelton 1988). The plant resistance may be caused by antixenosis, antibiosis or tolerance or a combination of these (Painter 1951, Kogan and Ortman 1978). The combination of these mechanisms can increase the effective life of an insect resistant cultivar and decrease the likelihood of a pest to overcome the resistance compared with any one mechanism especially in the case of antixenosis and antibiosis, as long as alternative sources of preferred hosts are available (Gould 1984). Breeding for genetic resistance against aphids has not been possible owing to the non-availability of resistance source within the crossable germplasms and lack of knowledge of the genetics of the trait (Bhatia *et al.* 2011). Therefore, the present study was undertaken to evaluate various genotypes of rapeseed-mustard to find out source(s) of resistance against mustard aphid and to identify the biochemical basis of resistance, if any.

## MATERIAL AND METHODS

### Field Studies

The present study was carried out at the Research Farm and Plant Protection laboratory, Department of Plant Breeding and Genetics, Punjab Agricultural University, Ludhiana (30.9º N, 75.85º E and 244 m above msl), India. Ten rapeseed-mustard genotypes were sown in a randomized split plot design in the 4th week of October. There were three replications with a plot size of 4 x 3 m each. The different genotypes were: *Brassica juncea*: RK 9501, Purple Mutant, RH 9501, RH 7846, JMM 927, Teri (00) R 9301; *B. napus* : Hyola PAC 401 hybrid in 2007-08; *B. carinata*: DLSC 2; *Eruca sativa*: T 27; *B. rapa* var. yellow *sarson*: YST 151 and *B. rapa* var. brown *sarson*: BSH 1 (susceptible check). These genotypes were sown in two different sets viz. protected and unprotected. The protected set was sprayed with oxydemeton methyl 25 EC @ 1000 ml ha$^{-1}$ when the *L. erysimi* population reached Economic Threshold Level (ETL) of 50-60 aphids/plant, while the unprotected set was left unsprayed. Standard agronomic practices were followed for raising a good crop except for spray of insecticides in unprotected set. The eggs and early instar larvae of large white butterfly, *Pieris brassicae* (Linnaeus) (Lepidoptera: Pieridae) were collected manually and destroyed. The data on population build up of *L. erysimi* were collected at weekly intervals to study the relative population build up on each genotype. For this, terminal 10 cm portion of the central shoot was examined to record the number of *alates* and *apterae*. In each plot, 10 such twigs were examined.

### Biochemical Constituents

At full bloom stage, top 10 cm portion was sampled from 10 plants/ plot for analyzing the biochemical constituents to find out relationship between aphid population and biochemical constituents, if any. Different biochemical constituents were determined using known standard methods i.e. glucosinolates by McGhee *et al.* (1965), total phenols by Swain and Hillis (1959), Ortho-dihydroxy phenols by Nair and Vaidyanathan (1964) and flavonols by Balbaa *et al.* (1974).

### Screen House Studies

The experiment was conducted in an insect-proof screen house of 30 mesh size at the Entomological Research Farm, Punjab Agricultural University, Ludhiana. Seeds of all the above mentioned genotypes were sown in earthen pots. Three seeds per pot were sown and after germination one seedling was retained. The experiment was set up in completely randomized design with three and five replications in 2006-07 and 2007-08, respectively. At 6 leaf stage, 20 apterous aphids were confined on each plant with the help of soft camel's hair brush following Bakhetia and Bindra

**Table 1. Relative population (mean ± SE) of *Lipaphis erysimi* on different *Brassica* genotypes/ hybrid under protected and unprotected conditions during 2006-07 and 2007-08 at Ludhiana**

| *Brassica* sp. | Geno-type/ Hybrid | 2006-07 | | 2007-08 | | Pooled | |
|---|---|---|---|---|---|---|---|
| | | Protect-ed | Unpro-tected | Protect-ed | Unpro-tected | Protect-ed | Unpro-tected |
| *Brassica juncea* | RK 9501 | 0.0 ± 0.0 | 15.9±0.9 | 1.1 ± 0.5 | 80.8 ± 2.0 | 0.5 ± 0.2 | 48.4 ± 1.5 |
| | RH 9501 | 3.1 ± 0.3 | 15.6 ± 0.3 | 4.3 ± 1.0 | 67.8 ± 7.4 | 3.8 ± 0.4 | 41.7 ± 3.6 |
| | RH 7846 | 4.2 ± 0.3 | 21.4 ± 3.5 | 1.9 ± 0.8 | 65.5 ± 5.8 | 3.1 ± 0.6 | 43.5 ± 1.3 |
| | JMM 927 | 0.0 ± 0.0 | 18.4 ± 0.7 | 1.5 ± 0.6 | 81.5 ± 8.4 | 0.8 ± 0.3 | 50.0 ± 3.9 |
| | Purple Mutant | 5.2 ± 0.5 | 20.7 ± 2.3 | 0.5 ± 0.5 | 62.8 ± 19.02 | 2.9 ± 0.5 | 41.7 ± 9.7 |
| | Teri (00) R 9903 | 5.3 ± 1.4 | 18.4 ± 1.3 | -- | -- | -- | -- |
| *B. napus* Hybrid | Hyola PAC 401 | -- | -- | 0.8 ± 0.2 | 78.0 ± 34.6 | -- | -- |
| *B. cari-nata* | DLSC 2 | 5.3 ± 2.3 | 16.3 ± 1.7 | 1.3 ± 0.5 | 25.5 ± 3.5 | 3.3 ± 1.3 | 20.9 ± 2.1 |
| *Eruca sativa* | T 27 | 1.0 ± 0.5 | 6.9 ± 1.4 | 0.0 ± 0.0 | 2.5 ± 0.5 | 0.5 ± 0.2 | 4.7 ± 0.5 |
| *B. rapa* var. yel-low *sar-son* | YST 151 | 3.4 ± 0.1 | 32.7 ± 2.1 | 2.9 ± 0.4 | 72.0 ± 4.0 | 3.2 ± 0.3 | 52.3 ± 2.8 |
| *B. rapa* var. brown *sarson* | BSH 1 | 2.5 ± 0.9 | 31.7 ± 1.7 | 6.3 ± 1.6 | 75.6 ± 4.7 | 4.4 ± 1.2 | 53.7 ± 1.8 |
| LSD (p<0.05) | | | | | | | |
| Protection | | | 1.3 | | 8.6 | | 4.2 |
| Genotypes | | | 2.9 | | 19.3 | | 9.2 |
| Interaction | | | 4.2 | | 27.3 | | 13.4 |

The protected set was sprayed with oxydemeton methyl 25 EC @ 1000 ml ha$^{-1}$.

(1977). The weekly observations on the number of offsprings produced by aphids and seedling mortality, if any, were recorded.

**Statistical analysis**

The data so obtained were subjected to analysis of variance as factorial experiments with randomized split plot design for field experiments and completely randomized design for screen house experiment (Snedecor and Cochran 1980). Means showing significant difference (p<0.05) were separated using Least Significant Difference.

## RESULTS AND DISCUSSION

**Field Studies**

In 2006-07 crop season, *L. erysimi* population ranged from 6.9 to 32.6 aphids/ plant on different gen-otypes in the unprotected set (Table 1). The maximum population of 32.6 aphids/ plant was observed on YST 151. It was followed by BSH 1 (31.7 aphids/ plant), which was *at par* with YST 151 but significantly (p<0.05) higher than rest of the genotypes. On the other hand, minimum population of 6.9 aphids/ plant was observed in T 27 which was significantly lower than the remaining genotypes.

In 2007-08 crop season, the aphid population ranged from 2.5 to 81.5 aphids/ plant on different genotypes/hybrid in the unprotected set. The maximum population (81.5 aphids/ plant) was observed in JMM 927 followed by RK 9501 (80.8 aphids/ plant) which were *at par* with each other. Population in these two genotypes was significantly higher than that in rest of the genotypes/ hybrid. On the other hand, minimum population of 2.5 aphids/ plant was observed on T 27

**Table 2.** Yield of different *Brassica* genotypes/ hybrids under protected and unprotected conditions during 2006–07 and 2007-08 at Ludhiana and the resultant yield losses

| *Brassica* sp. | Genotype/ Hybrid | 2006–07 | | | | 2007-08 | | | | Pooled | | | |
|---|---|---|---|---|---|---|---|---|---|---|---|---|---|
| | | P* | UP | Mean | Yield Loss (%) | P | UP | Mean | Yield Loss (%) | P | UP | Mean | Yield Loss (%) |
| *Brassica juncea* | RK 9501 | 1865 ±79 | 1398 ±20 | 1631.5 ±48.6 | 25.04 | 2041 ±146 | 1533 ±25 | 1787.0 ±85.8 | 24.88 | 1953.0 ±50.8 | 1465.5 ±4.7 | 1709.2 ±27.5 | 24.96 |
| | RH 9501 | 1811 ±119 | 1495 ±25 | 1653.0 ±55.7 | 17.44 | 2243 ±92 | 1590 ±186 | 1916.5 ±88.6 | 29.11 | 2027.0 ±17.1 | 1542.5 ±80.8 | 1784.7 ±48.2 | 23.90 |
| | RH 7846 | 1612 ±192 | 1448 ±65 | 1530.0 ±125.7 | 10.17 | 1694 ±137 | 1444 ±50 | 1569.0 ±72.5 | 14.75 | 1653.0 ±113.3 | 1446.0 ±17.5 | 1549.5 ±53.8 | 12.52 |
| | JMM 927 | 1775 ±18 | 1474 ±25 | 1624.5 ±20.5 | 16.95 | 1659 ±183 | 1347 ±174 | 1503.0 ±170.1 | 18.80 | 1717.0 ±83.0 | 1410.5 ±75.1 | 1563.7 ±75.2 | 17.85 |
| | Purple Mutant | 858 ±17 | 611 ±20 | 734.5 ±16.1 | 28.78 | 977 ±95 | 563 ±78 | 770.0 ±80.4 | 42.37 | 918.0 ±52.7 | 587.0 ±46.4 | 752.5 ±48.27 | 36.05 |
| | Teri (00) R 9903 | 1180 ±41 | 1029 ±11 | 1104.5 ±15.0 | 12.80 | — | — | — | — | — | — | — | — |
| *B. napus* Hybrid | Hyola PAC 401 | — | — | — | — | 2104 ±175 | 1729 ±136 | 1916.5 ±150.2 | 17.82 | — | — | — | — |
| *B. carinata* | DLSC 2 | 1061 ±43 | 980 ±55 | 1020.5 ±14.0 | 7.63 | 1091 ±59 | 944 ±50 | 1017.5 ±46.2 | 13.47 | 1076 ±25.1 | 962.0 ±13.4 | 1019.0 ±16.4 | 10.59 |
| *Eruca sativa* | T 27 | 126 ±9 | 119 ±6 | 122.5 ±3.9 | 5.55 | 202 ±37 | 190 ±5 | 196.0 ±18.6 | 5.94 | 164.0 ±19.5 | 154.5 ±4.9 | 159.2 ±10.6 | 5.79 |
| *B. rapa* var. yellow sarson | YST 151 | 506 ±65 | 452 ±48 | 479.0 ±50.3 | 10.67 | 805 ±100 | 666 ±134 | 735.5 ±116.8 | 17.26 | 655.5 ±79.6 | 559.0 ±75.0 | 607.2 ±76.7 | 14.72 |
| *B. rapa* var. brown sarson | BSH 1 | 746 ±63 | 619 ±51 | 682.5 ±31.5 | 17.02 | 979 ±98 | 798 ±54 | 888.5 ±75.7 | 18.48 | 862.5 ±58.8 | 708.5 ±52.7 | 785.5 ±50.0 | 17.86 |
| | Mean | 1154.0 ±21.7 | 962.5 ±5.4 | 1058.3 ±11.7 | | 1379.5 ±10.7 | 1080.4 ±51.4 | 1229.9 ±30.8 | | 1225.1 ±11.15 | 981.7 ±19.70 | 1103.4 ±14.4 | |
| | LSD (p=0.05) Protection Genotypes Interaction | 59.17 132.32 187.13 | | | | 102.42 229.01 NS | | | | 54.79 122.53 173.28 | | | |

*P: Protected, UP: Unprotected
The protected set was sprayed with oxydemeton methyl 25 EC @ 1000 ml ha⁻¹.

Table 3. Biochemical constituents in relation to aphid population in inflorescence part of various *Brassica* genotypes/ hybrid

| *Brassica* sp. | Genotype/ Hybrid | Total phenols (mg/g) | | | Ortho-dihydroxy phenols (mg/g) | | | Flavonols (mg/g) | | | Glucosinolates (µ mole/g) | | | Aphid population/ plant | | |
|---|---|---|---|---|---|---|---|---|---|---|---|---|---|---|---|---|
| | | 2006-07 | 2007-08 | Pooled Mean | 2006-07 | 2007-08 | Pooled Mean | 2006-07 | 2007-08 | Pooled Mean | 2006-07 | 2007-08 | Pooled Mean | 2006-07 | 2007-08 | Pooled Mean |
| *Brassica juncea* | RK 9501 | 5.63 | 8.18 | 6.91 | 1.01 | 0.64 | 0.83 | 4.66 | 1.45 | 3.06 | 45.49 | 72.83 | 59.16 | 15.9 | 80.8 | 48.4 |
| | RH 9501 | 7.00 | 9.54 | 8.27 | 0.94 | 0.56 | 0.76 | 2.80 | 1.45 | 2.13 | 64.60 | 92.94 | 78.77 | 15.6 | 67.8 | 41.7 |
| | RH 7846 | 8.91 | 8.18 | 8.55 | 0.45 | 0.42 | 0.44 | 2.13 | 0.73 | 1.43 | 45.16 | 76.94 | 61.05 | 21.4 | 65.5 | 43.5 |
| | JMM 927 | 6.27 | 8.64 | 7.46 | 0.90 | 0.56 | 0.73 | 3.99 | 1.1 | 2.55 | 40.33 | 64.94 | 52.64 | 18.4 | 81.5 | 50.0 |
| | Purple Mutant | 7.63 | 13.18 | 10.41 | 1.01 | 0.92 | 0.97 | 4.66 | 2.4 | 3.53 | 70.66 | 108.43 | 89.55 | 20.7 | 62.8 | 41.7 |
| | Terri(00) R 9903 | 7.72 | -- | -- | 1.01 | -- | -- | 3.86 | -- | -- | 49.04 | -- | -- | 18.4 | -- | -- |
| *B. napus* Hybrid | Hyola PAC 401 | -- | 9.99 | -- | -- | 0.66 | -- | -- | 1.45 | -- | -- | 15.49 | -- | -- | 78.0 | -- |
| *B. carinata* | DLSC 2 | 8.36 | 14.54 | 11.45 | 1.13 | 0.99 | 1.06 | 4.26 | 2.55 | 3.41 | 77.74 | 110.78 | 94.26 | 16.3 | 25.5 | 20.9 |
| *Eruca sativa* | T 27 | 8.91 | 11.36 | 10.14 | 2.06 | 0.99 | 1.53 | 2.13 | 1.1 | 1.62 | 98.70 | 126.74 | 112.72 | 6.9 | 2.5 | 4.7 |
| *B. rapa* var. yellow sarson | YST 151 | 5.27 | 9.99 | 7.63 | 0.79 | 0.75 | 0.77 | 3.33 | 1.64 | 2.49 | 46.45 | 82.73 | 64.59 | 32.7 | 72.0 | 52.3 |
| *B. rapa* var. brown sarson | BSH 1 | 7.09 | 9.09 | 8.09 | 0.90 | 0.71 | 0.81 | 4.80 | 1.45 | 3.13 | 47.82 | 80.47 | 64.15 | 31.7 | 75.6 | 53.7 |
| **Correlation Coefficient** | | -0.47 | -0.64 | -0.75 | -0.66 | -0.73 | -0.82 | 0.35 | -0.19 | 0.25 | -0.63 | -0.72 | -0.90 | | | |

$R^2$: 2006-07=0.53, 2007-08=0.92, Pooled=0.94

being significantly lower than the remaining genotypes/hybrid. In the case of DLSC 2, the aphid population was slightly higher than that in T 27, it was significantly lower than the remaining genotypes/hybrid.

From the two years pooled data it is evident that the maximum population of 53.7 aphids/ plant was observed in BSH 1, which was significantly higher than the remaining genotypes, except YST 151 (52.3 aphids/plant). On the other hand, aphid population in T 27 (4.7 aphids/ plant) and DLSC 2 (20.9 aphids/ plant) was significantly lower than the remaining genotypes.

## Yield Performance

### Crop Season 2006-07

The mean seed yield of 962.5 kg ha$^{-1}$ was recorded in unprotected plots compared to 1154.0 kg ha$^{-1}$ in protected plots (Table 2). Different genotypes differed significantly ($p<0.05$) with respect to seed yield. The maximum seed yield (1653 kg/ha) was observed in RH 9501. It was followed by RK 9501 (1631.5 kg), JMM 927 (1624.5) and RH 7846 (1530 kg). Seed yield in these four genotypes was on a par with each other and significantly higher than the remaining genotypes. On the other hand, minimum seed yield (122.5 kg) was recorded in T 27. Seed yield in YST 151 and T 27 was significantly lower than the remaining genotypes.

The estimated avoidable losses in seed yield due to *L. erysimi* infestation ranged from 5.55 to 28.78 per cent in different genotypes. Purple mutant suffered the maximum (28.78%) loss in seed yield. It was followed by RK 9501 (25.04%), RH 9501 (17.44%), BSH 1 (17.02), JMM 927 (16.95), Teri (00) R 9903 (12.80%), YST 151 (10.67%), RH 7846 (10.17%), DLSC 2 (7.63%) and T 27 (5.55%).

### Crop Season 2007-08

The mean seed yield of 1080.4 kg ha$^{-1}$ was recorded in unprotected set compared to 1379.5 kg in protected set. The maximum yield (1916 kg/ha) was recorded in RH 9501 and *B. napus* hybrid Hyola PAC 401. It was followed by RK 9501 (1787 kg/ha). The seed yield in these two genotypes and one hybrid was on a par with each other but significantly higher than the remaining genotypes. The minimum seed yield (196 kg/ha) was recorded in T 27, which was significantly lower than all the other genotypes/hybrid.

The loss in seed yield ranged from 5.94 to 42.37 per cent in different genotypes/hybrid. Maximum loss (42.37%) was recorded in Purple Mutant. It was followed by RH 9501 (29.11), RK 9501 (24.88), BSH 1 (18.48), JMM 927 (18.80), Hyola PAC 401 (17.82), YST 151 (17.26), RH 7846 (14.75), DLSC 2 (13.47) and T 27 (5.94).

### Pooled

The pooled data on seed yield of both the years revealed that mean yield of 981.7 kg ha$^{-1}$ was recorded in unprotected plots as against 1225.1 kg in protected ones. RH 9501 recorded the maximum yield (1784.7 kg/ha) followed by RK 9501 (1709.2kg), which were *at par* with each other but significantly higher than rest of the genotypes. The seed yield in T 27 (159.5kg) and YST 151 (607.2kg) was significantly lower than the remaining genotypes.

The loss in seed yield varied from 5.79 to 36.05 per cent in different genotypes. Purple Mutant suffered the maximum loss in yield (36.05%). It was followed by RK 9501 (24.96%), RH 9501 (23.90), BSH 1 (17.86), JMM 927 (17.85), YST 151 (14.72), RH 7846 (12.52), DLSC 2 (10.59) and T 27 (5.79). During both the years of study, T 27 and DLSC 2 suffered the least pest damage.

## Biochemical Constituents in Relation to Aphid Infestation

The various biochemical constituents *viz.* total phenols, ortho-dihydroxy phenols and glucosinolates showed a significant negative correlation with aphid population during both the years of study (Table 3), whereas, flavonols did not exhibit any significant correlation with aphid population. Analysis of combined effect of biochemical constituents showed that over 94 per cent variation in aphid population ($R^2$=0.94) can be ascribed to combined effect of these biochemical constituents studied. While comparing the role of individual component in reducing aphid population in the field, it is surmised that glucosinolates, ortho-dihydroxy phenols and total phenols are important in that order.

## Screen House Studies

In screen house trial, an increase in aphid population was observed up to 21 days after their release on the seedlings with peak population during 21 and 14 days after release in most of the genotypes in 2006-07 and 2007-08, respectively.

In 2006-07, the significant ($p<0.05$) differences

Table 4. Seedling survival of oilseed *Brassica* genotypes after artificial aphid infestation (Screen house studies)

| Brassica sp. | Genotype | Population initially released | Aphid population/ plant (Days after release) | | | | | | | | Seedling mortality after 30 days (%) | |
|---|---|---|---|---|---|---|---|---|---|---|---|---|
| | | | 7 | | 14 | | 21 | | 28 | | | |
| | | | 2006-07 | 2007-08 | 2006-07 | 2007-08 | 2006-07 | 2007-08 | 2006-07 | 2007-08 | 2006-07 | 2007-08 |
| *Brassica juncea* | RK 9501 | 20 | 26.67 ±8.82* | 67.00 ±11.14** | 35.00 ±7.64 | 52.00 ±14.63 | 23.33 ±8.82 | 39.00 ±9.27 | 16.67 ±8.82 | 32.00± 8.75 | 33.3 | 0.0 |
| | RH 9501 | 20 | 43.33 ±3.33 | 105.00 ±9.75 | 30.00 ±11.55 | 70.00 ±8.37 | 20.00 ±0.00 | 20.00 ±5.24 | 16.67 ±1.67 | 15.00± 5.61 | 33.3 | 20.0 |
| | RH 7846 | 20 | 40.00 ±15.28 | 56.00 ±4.30 | 30.00 ±15.28 | 25.00 ±4.47 | 26.67 ±12.02 | 19.60 ±3.41 | 11.67 ±4.41 | 11.00± 3.67 | 0.0 | 0.0 |
| | JMM 927 | 20 | 56.67 ±21.86 | 33.00 ±3.74 | 46.67 ±13.33 | 78.00 ±18.82 | 51.67 ±9.28 | 36.67 ±11.14 | 26.67 ±13.33 | 25.00± 7.42 | 33.3 | 40.0 |
| | Purple Mutant | 20 | 26.67 ±8.82 | 32.00 ±3.39 | 36.67 ±6.67 | 44.00 ±3.32 | 55.00 ±18.56 | 94.00 ±6.40 | 35.00 ±12.02 | 30.00± 8.46 | 33.3 | 40.0 |
| | Terri(00) R 9903 | 20 | 56.67 ±6.67 | -- | 66.67 ±8.82 | -- | 75.00 ±25.17 | -- | 45.00 ±15.28 | -- | 33.3 | -- |
| *B. napus* Hybrid | Hyola PAC 401 | 20 | -- | 113.00 ±25.28 | -- | 212.50 ±53.85 | -- | 55.00 ±11.66 | -- | 43.00± 9.84 | -- | 20.0 |
| *B. carinata* | DLSC 2 | 20 | 33.33 ±8.82 | 88.00 ±17.72 | 41.67 ±14.24 | 140.00 ±40.50 | 15.00 ±2.89 | 87.50 ±25.50 | 13.33 ±3.33 | 43.75± 12.35 | 0.0 | 20.0 |
| *Eruca sativa* | T 27 | 20 | 20.00 ±10.00 | 60.00 ±8.37 | 26.67 ±8.82 | 37.00 ±7.00 | 10.00 ±5.77 | 10.00 ±2.74 | 0.00 ±0.00 | 0.00±0 .00 | 0.0 | 0.0 |
| *B. rapa* var. yellow sarson | YST 151 | 20 | 46.67 ±12.02 | 48.00 ±9.70 | 65.00 ±29.63 | 64.00 ±13.27 | 30.00 ±10.00 | 18.75 ±8.94 | 10.00 ±3.33 | 7.50±2 .00 | 66.7 | 60.0 |
| *B. rapa* var. brown sarson | BSH 1 | 20 | 26.67 ±3.33 | 172.00 ±8.60 | 70.00 ±20.82 | 228.00 ±14.63 | 106.67 ±34.80 | 144.00 ±9.27 | 30.00 ±10.00 | 55.00± 13.57 | 66.7 | 60.0 |
| CD(p<0.05) | | | NS | 34.59 | NS | 68.18 | 48.96 | 32.07 | NS | 23.75 | | |

*Mean of 3 replications  **Mean of 5 replications

in aphid population on different genotypes were observed only after 21 days of release. The maximum population of aphids (106.67/plant) during this period was observed on BSH 1 followed by Teri (00) R 9903 (75.00 aphids/plant) which were *at par* with each other, but significantly higher than the remaining genotypes. The genotypes T 27 and DLSC 2 harboured the lowest aphid population i.e. 10.00 and 15.00 aphids/ plant, respectively. The maximum seedling mortality of 66.7 per cent was recorded in genotypes BSH 1 and YST 151, each, whereas T 27, DLSC 2 and RH 7846 did not suffer any seedling mortality.

In 2007-08, the maximum aphid population (228.00/ plant) was observed on BSH 1, 14 days after release. It was followed by Hyola PAC 401 (212.50 aphids/ plant) and DLSC 2 (140.00 aphids/ plant). Population in these two genotypes and one hybrid was on a par with each other but significantly higher than the remaining genotypes. On the other hand, genotypes RH 7846 and T 27 harboured the minimum aphid population i.e. 25.00 and 37.00 aphids/plant, respectively.

The maximum seedling mortality of 60 per cent was recorded in genotypes BSH 1 and YST 151, each. On the other hand, genotypes T 27, RK 9501 and RH 7846 did not suffer any seedling mortality.

There was a considerable variation in the population of *L. erysimi* on different *Brassica* genotypes and the resultant yield losses. During both the years of study significantly low population was recorded on T 27 and DLSC 2, which was probably due to poor olfactory stimulus offered by these to the initial alate settlers (Harish Chander *et al.* 1997). Since, *L. erysimi* is a specialist feeder of oilseed *Brassica*, it is expected to utilize concentration of glucosinolates in T27 and DLSC2 as cues to locate its host plant. Thus, high population levels of this pest on these two genotypes are expected. But actually it was the opposite. It may possibly be due to the complex role of various other primary and secondary plant chemicals (van Dam and Oomen 2008, Hopkins *et al.* 2009) as is evident from the high concentration of ortho-hydroxy and total phenols in these genotypes in the present study. On the other hand, BSH 1 and YST 151 i.e. *B. rapa* group harboured quite high aphid population and suffered high yield losses and seedling mortality. *B. juncea* group, in general, suffered comparatively low seed yield losses and seedling mortality while harbouring considerable aphid population, except Purple Mutant. This can be ascribed to inherent genetic characters of *B. rapa* and *B. juncea* group, respectively (Pathak 1961, Bindra and Deole 1962, Singh *et al.* 1965, Kundu and Pant 1967, Kher and Rataul 1992 a, b). The high yield losses in Purple Mutant consequent to aphid infestation had also been reported earlier (Anonymous 2006).

The observed negative correlation of various biochemical constituents with aphid population is in conformity with the findings of earlier workers such as phenols in *Eruca sativa* (Narang 1982), phenols in different *Brassica* genotypes (Sachan and Sachan 1991), glucosinolates in *B. napus* (Gill and Bakhetia 1985), glucosinolates in different *Brassica* species (Bakhetia *et al.* 1982, Malik 1981, Singh *et al.* 2000). Dilawari and Atwal (1987) observed that the number of probes increased and feed uptake reduced significantly in an artificial media containing higher level of glucosinolates. High antibiosis in T 27 due to presence of phenols and glucosinolates was observed by Bakhetia and Bindra (1977) and in different genotypes of *E. sativa* by Kundu and Pant (1967) and Narang (1982).

## CONCLUSION

Thus, it can be concluded from this study that BSH 1 and YST 151 suffered high yield loss due to aphid infestation. On the other hand, T 27 and DLSC 2 harboured significantly lower aphid population and suffered the least yield loss and seedling mortality and can serve as important sources of resistance against mustard aphid in breeding programmes aimed at developing mustard/turnip aphid resistant cultivars.